\documentstyle[epsfig]{mn}

\input{psfig.sty}

\title[SMBH and the downsizing of massive ellipticals]
{Supermassive black holes, star formation and
downsizing of elliptical galaxies}

\author[A. Pipino, J. Silk and F. Matteucci]{Antonio Pipino$^{1,2}$, 
Joseph Silk$^1$ and Francesca Matteucci,$^3$\\
$^1$Astrophysics, University of Oxford, Denys Wilkinson Building,
    Keble Road, Oxford, OX1 3RH, U.K.\\
$^2$Department of Physics and Astronomy, University of Southern California, 
Los Angeles, CA 90089-0484 \\
$^3$Dipartimento di Astronomia, Universit\'a di Trieste,
    Via G.B. Tiepolo, 11, I-34127, Trieste, Italy}
\date{Accepted,
      Received }

\begin{document}
\maketitle

\begin{abstract}

The overabundance of Mg 
relative to Fe, observed in the nuclei of bright ellipticals, and its increase with galactic mass, poses a serious problem for all 
current models of galaxy formation.
Here we  improve on  the { one-zone} chemical evolution models for 
elliptical galaxies by taking into account 
positive feedback produced in the early stages of  super-massive central black hole growth.
We can account for both the observed correlation and the scatter if the observed anti-hierarchical behaviour of the AGN population couples to  galaxy assembly 
and results in an enhancement of  the star formation efficiency which is proportional to  galactic mass.
{ At low and intermediate galactic masses, however, a slower mode for star formation suffices
to account for the observational properties.}

\end{abstract}

\begin{keywords}
galaxies: ellipticals: chemical abundances, formation and evolution
\end{keywords}

\section{Introduction}

Increasing evidence  has accumulated over the past decade
that  the [Mg/Fe] ratio is super-solar
in the cores of bright galaxies (e.g. Faber et al., 1992; Carollo et
al., 1993). An overabundance of Mg relative 
to Fe is the key indicator  that galaxy formation occurred before  a substantial number 
of type Ia SNe could explode and contribute to lower the [Mg/Fe] ratio (for the time-delay model, see Matteucci 
2001). In addition, the [Mg/Fe] ratio  in the cores of ellipticals increases with galactic mass 
(Worthey et al. 1992; Weiss et al. 1995; Kuntschner 2000, Kuntschner et al. 2001, Thomas et al. 2005, Nelan et al. 2006).
This relation seems to already  be in place  at redshift 0.4 (Ziegler et al 2006).

In order to account for this trend in star formation
time-scale, and specifically of  
increase in  the star formation efficiency with galactic mass, there are at least three 
possibilities that have been discussed in the literature.
One involves  loss of the residual gas via galactic winds
 that are initiated
earlier in the most massive objects (inverse wind picture, see Matteucci 1994).
Another possibility is to assume that the initial mass function (IMF) systematically 
becoming flatter with increasing galactic mass (e.g., as argued by van Dokkum 2008).
A top-heavy IMF was applied by Nagashima et al. (2005) to reproduce the mass-[Mg/Fe] relation,
although this attempt proved to be unsuccessful.
A  selective loss of metals could also be the cause 
for the increase of [Mg/Fe] (see Matteucci et al. 1998).
We consider that the first possibility is the best motivated, 
and we will refer to it as  \emph{chemical downsizing}. Observational evidence for chemical downsizing from $z\sim 3$
has recently been obtained by Maiolino et al. (2008).

\emph{Chemical downsizing} was predicted neither by the 'classical wind 
scenario' of Larson (1974), where in bigger galaxies (i.e. deeper potential wells) the star formation time-scale is longer
than in low mass counterparts, nor by hierarchical assembly in the CDM scenario (White and Rees 1978; Kauffmann \& Charlot 1998), where the spheroids with
the largest masses are the last ones to be formed (accordingly, in  the \emph{time-delay} model, they suffer
strong injection of Fe from SNIa). The so-called revised
 \emph{monolithic} scenario
(Matteucci, 1994; Chiosi \& Carraro 2002) does explain both the mass-[Mg/Fe] (MFMR, hereafter) and the mass-metallicity (MMR)
relations, but physical motivation is lacking. 
In particular,  Pipino \& Matteucci (2004, PM04 hereafter) used  the inverse
wind scenario (Matteucci 1994) plus an initial infall episode within a multi-zone
formulation to account  simultaneously for
the whole set of chemical and photometric observables. In contrast, 
a merger-induced star formation (SF) history  produces
results at variance with the MFMR (e.g. Thomas \& Kauffmann 1999, Pipino \& Matteucci, 2006, 2008).
However none of these models addressed the associated issue of mass downsizing, which clearly must go hand in hand with chemical downsizing.

Study of the evolution of  the galaxy luminosity function with redshift (Bundy et al. 2006, Scarlata et al 2006; Perez-Gonzalez
et al. 2007) has demonstrated \emph{mass downsizing}, 
in the sense that the stellar mass of the most massive spheroids
seem to be passively evolving from high redshift.
At the same time, the active galactic nuclei (AGN) population behaves  anti-hierarchically (e.g. Hasinger et al. 2005).
Substantial modification of the mass assembly of stars relative to assembly of baryons, at least
with respect to CDM, seems to be required.

It is interesting to note that the peak of luminous AGN activity occurs around redshift 2 where
the SF activity also seems to peak. In fact there is a time delay for low luminosity AGN; these peak several Gyr after the peak in SF activity.
Moreover, the correlation between central black hole mass
and spheroid velocity dispersion  (Gebhardt et al. 2000; Ferrarese and Merritt 2000) suggests that the black holes
form contemporaneously with the spheroids (Dietrich and Hamann 2004).
The gas-rich protogalaxy provides the ideal accretion environment for forming the super-massive
black hole (SMBH).  
There is indeed a natural coupling between the two processes, since the SMBH
undergoes most of its growth in the gas-rich phase and the SMBH outflow pressurises the gas,
which in turn forms stars. The nature, and indeed the direction, of the 
triggering is unclear, and it is worth mentioning that hitherto only models
which incorporate negative (namely which quenches SF) feedback have been simulated.
By means of different recipes, these models aim at reproducing the above mentioned
observations. 
From the point of view of the cosmological simulations, 
AGN quenching of the cooling - and
thus of the star formation -  in the
\emph{QSO mode} is required to truncate star formation in early-type galaxies (ETGs), and AGN activity   in the \emph{radio mode}  (of outflow) (Silk \& Rees, 1998)
is required to suppress infall and late star formation and thereby maintain the red colours of ETGs.
 
To-date, however, none of these cosmologically-motivated models have succeeded
in reproducing the chemical downsizing inferred from the current epoch data. The disagreement with observations
seems to worsen when the time evolution of the MMR is taken
into account (Maiolino et al. 2008).
The monolithic approach to chemical evolution models for single galaxies 
does allow one to reproduce the data.
For example, Romano et al (2002) implemented
a recipe for quenching  SF earlier in the most massive galaxies by means of an AGN.
At the same time, Kobayashi et al. (2007) managed to reproduce the observed trend by means of SNe and hypernovae,
following PM04.
No one, however, has so far studied  the effects of  SMBH-triggered star formation
in a self-consistent way that is applicable to cosmological simulations.
 In particular, the recipes for star formation need to be addressed.

The aim of the present paper is to take a first step towards addressing the role
of positive (i.e. which boosts SF) SMBH feedback in numerical simulations of galactic chemical evolution.
A chemical evolution model for a single galaxy is the ideal tool
for  studying the reliability of such an approach, because the results
can be directly compared with the observed MFMR.
In particular, we implement the star formation modes suggested by Silk
(2005) into the PM04 model for the chemical evolution of ETGs.
According to Silk (2005), star formation is triggered coherently and rapidly
by a SMBH  jet-induced hot plasma cocoon which over-pressures cold
clouds and induces collapse within the central core of the forming
galaxy. This is one of the  consequences  of the propagation of
a broad jet through an inhomogeneous
 interstellar medium with a low dense cloud filling factor (e.g., Saxton et al. 2005;
Antonuccio-Delogu \&  Silk 2008).
{ Krause \& Alexander (2007)
present simulations of the Kelvin-Helmhotz instability 
with clouds of differing density. The clouds are shocked, collapse into a filament, and then disperse into cloudlets and 
more filaments. Over time, more gas condenses 
into the cold phase. Mellema et al. (2002) then see  fragmentation 
into small stable cloudlets, which may harbour star formation (Fragile et al. 2004).}
The feedback is positive, as there is insufficient time for the 
supernova-driven negative feedback to develop.
 The interaction of the outflow with the surrounding
protogalactic gas at first stimulates star formation on a short time-scale.
Evidence has been found
for jet-stimulated star formation up to $z\sim 4-5$ ({ Bicknell et al. 2000}, Venemans et
al. 2004), followed in at least one case by a starburst-driven superwind
(Zirm et al. 2005). There is also evidence for triggering of molecular gas at high redshift (Klamer et al. 2004).
{ The prototypical low-redshift example is  Minkowski's object (van Breugel et al. 1985, Croft et al. 2006),
where the neutral hydrogen may have cooled out of a warmer, clumpy intergalactic or interstellar medium as a result of jet interaction, 
followed by collapse of the cooling clouds and subsequent star formation, unlike the jet-induced star formation in Centaurus A, where the jet interacts with pre-existing cold gas.
Another interesting case is the spiral galaxy
NGC 4258 which exhibits a concentration of CO along the jets that  is similar to what is expected as fuel for jet-induced 
star formation in more distant objects (e.g., Krause et al. 2007). We finally note that there is 
evidence of a higher type Ia supernova rate in radio-loud ellipticals (Della Valle et al. 2005)
that may be connected to a small amount of recent (jet-induced) star formation (Antonuccio-Delogu \& Silk, in preparation).}

We  first review the main ingredients of our chemical evolution models
and present the new modifications in Sec. 2.
Results and conclusions will be presented in Secs. 3 and 4, respectively

\section{The model}

The adopted chemical evolution model
is an updated version of the multi-zone model of Pipino \& Matteucci (2004). 
{ Note that, in order to compare our results with the latest available
data by Thomas et al. (2008), which pertain to the whole galaxies
rather than only to their central parts, we will run the model
in a one-zone fashion out to one effective radius. 
This assumption is also required because we want to take into
account the physics of the jet (see below) without 
detailed modelling of the gas transfer between the shells.}
We calculate the evolution of the elemental abundances
by means of the equation of chemical evolution:
\begin{eqnarray*}
{d G_i (t) \over d t}  =  -\psi (t) X_i (t)\, + \\
                         +\int_{M_L}^{M_{B_m}} \psi (t-\tau_m) Q_{mi}(t-\tau_m) \phi (m) dm\, +\\
                         + A\int_{M_{B_m}}^{M_{B_M}} \phi (m) \left [ \int_{\mu_{min}}^{0.5} f(\mu) \psi (t-\tau_{m_2}) Q_{mi} (t-\tau_{m_2}) d\mu \right ] dm \, +\\
                         +(1-A)\, \int_{M_{B_m}}^{M_{B_M}} \psi (t-\tau_m) Q_{mi}(t-\tau_m) \phi (m) dm\, +\\
                         +\int_{M_{B_M}}^{M_U} \psi (t-\tau_m) Q_{mi}(t-\tau_m) \phi (m) dm\,  +\\
		         +({d G_i (t) \over d t})_{infall}\,  ,	
\end{eqnarray*}
where $G_i (t)= \rho_{gas}(t) \,X_i (t)$ is the mass density
of the element \emph{i} at the time \emph{t}.
$X_i (t)$ 
is defined as the abundance by mass of each element \emph{i}.
By definition $\sum_i X_i=1$. We refer the reader
to Matteucci $\&$ Greggio (1986) 
for a comprehensive discussion of this equation. Here we note
that the integrals on the right-hand side of the equation give the 
rate at which the element \emph{i} is restored into the interstellar medium (ISM)
as unprocessed or newly-synthesized elements by low- and intermediate-mass 
stars, SNIa and SNII, respectively. The last term represents the infall rate,
namely the rate at which the gas is assembling to form the galaxy. 
$\phi (m)\propto m^{-(1+x)}$ is the IMF, normalized
to unity in the mass interval $0.1 -100 M_{\odot}$. In the following
we shall only use the exponent $x=1.35$ (Salpeter, 1955).

\subsection{The growth of a SMBH}

We deal with SMBH growth in a simple way,
namely we note that the time-scale for  star formation
set by chemical downsizing is of the order of $\sim 0.5$ Gyr
for the most massive galaxies (see PM04's best model results).
This is roughly the time needed for a SMBH to grow
from a seed of $\sim 10^3 M_{\odot}$ to the mass
required to fit the local Magorrian (1998) relation.
Therefore we will assume this to happen
without any further hypotheses.

We refer the reader to Silk (2005) for a more detailed analysis of 
the SMBH growth during the formation of a massive spheroid.
Here we note that one may crudely describe this situation by modelling the late-time
cocoon-driven outflow as quasi-spherical, and 
use  a spherical shell approximation
to describe the swept-up protogalactic gas. Initially, following
Begelman and Cioffi (1989), the interaction of the pair of jets may be
modelled by introducing an over-pressured and much larger cocoon, the ends of which
advance into the protogalactic gas at a speed $v_J$  and which expands
laterally at a speed determined eventually  by pressure balance with the ambient
gas. 

Only a small fraction of the protogalactic gas reservoir is implicated in AGN
feeding, even if this occurs at the maximum (Bondi) accretion rate.
Super-Eddington outflow  of course requires
super-Eddington  accretion which is plausibly associated with an
Eddington luminosity-limited luminous phase   implicated in the need to
generate massive SMBHs by $z\sim 6$ (Volonteri and Rees 2005).
A time-scale of $10^6-10^7$ yr for the super-Eddington phase would
more than suffice to provide the accelerated triggering of associated star formation.
The SMBH grows mostly in the super-Eddington phase while 
most of the spheroid stars  grow during the Eddington phase.  The latter 
phase ends by
quenching the feeding source, when the outflow clears out the remaining
gas. Only then is spheroid star formation terminated.

\subsection{Recipes for  star formation}

The variable $\psi$ in the equation of the chemical evolution is the star formation rate, for which PM04 adopted the following law: 
\begin{equation}
\psi (t)= \nu \cdot M_{gas} (t)\, ,
\end{equation} 
namely it is assumed to be proportional to the gas density
via a constant $\nu$ which represents the star formation efficiency.
We assume $\nu = \nu_{PM04}$, namely as an increasing function of the galactic mass 
(see Table 1) in order to reproduce the 'inverse wind
model' as originally suggested by Matteucci (1994) and to recover PM04's best model (their model II).  The star formation
history is thus determined by the interplay
between the infall time-scale at that radius, the star formation
efficiency and the occurrence of the galactic wind (i.e. the energetic
feedback from SNe and stellar winds).  In each zone, we assume that
$\psi=0$ after the development of the wind.
 
Silk (2005) proposes that there are two modes of star formation, writing $
\psi(t) =\alpha_{S,J}f_g\sigma_g^3/G,$ where 
$\alpha_{S}$ pertains to a slow self-regulated mode,
similar to the disks of spirals, whereas $\alpha_{J}$ takes into
account SF boosting by SMBH. $G$ is the constant of gravity, $f_g$ the gas fraction, $\sigma_g$ the gas velocity
dispersion.

{ In more detail, we have:
\begin{equation}
\alpha_{S}\sim 0.01 ({\sigma_g\over 10 km s^{-1}})
({v_{sh}\over 400 km s^{-1}}) ({10^{51} erg \over E_{SN}})( {m_{SN}\over 100 M_{\odot}}\, ,)
\end{equation}} 
where $v_{sh}$ the velocity of the SN remnant shell, , $E_{SN}$ the initial energy
of a SN explosion and $m_{SN} \sim 130 M_{\odot}$ is the mass of stars
needed per each SNII explosion for the Salpeter IMF.
{ An order of magnitude estimate give us $\alpha_{S}\sim$ 0.01-0.05 for low
mass ellipticals. 
whereas for more massive objects we expect $\alpha_{S}\sim$ 0.2.}
On the other hand, $\alpha_J$ is set 
by the super-Eddington outflow. 
{ Silk (2005) shows that in this case the star formation rate
can be written as $\psi (t)\sim L_{J}/cv_\infty  \sim 
 f_c\sigma_g v_wv_c^2/(v_JG)$, where $L_{J}$ is the jet luminosity, $v_\infty$ 
is the terminal velocity of the outflow,  
$f_c$ is the baryon compression in the galactic core, $v_w$ the wind velocity, $v_c$ 
is the cocoon expansion velocity; finally $v_J$ is the jet velocity.
Comparing the above expression for $\psi(t)$ to et. (1)
we derive:
\begin{equation}
\alpha_J \sim (f_c/f_g)(v_w/v_J) (v_c/ \sigma_g)^2
\sim \frac{L_J}{L_{cr}},
\end{equation}
namely $\alpha_J$ scales as the 
the jet luminosity in units of the critical luminosity $L_{cr}$
needed to expel all of the protogalactic gas.}

Silk (2005)'s conjecture is the following.
The outflow is super-Eddington until the cocoon is limited by  ambient
pressure and becomes quasi-spherical.
As the massive star formation/death rate
slows, the AGN feeding augments and the outflow stimulates more star
formation.  Therefore the net effect is that the star formation has negative feedback on AGN feeding,
whereas the AGN feeding has positive feedback on star formation. { At this stage we have $\alpha_{J} \sim \frac{L_J}{L_{cr}}\sim 1$. Clearly
the one-zone chemical evolution model cannot properly resolve the evolution of a jet and
its interaction with the surrounding medium. In order to avoid
the use of too many free parameters, we simply assume that the jet is playing a role in triggering SF by setting $\alpha_{J}=1$.}

We can rewrite the two modes suggested by Silk (2005) as:
\begin{equation}
\psi (t)= \nu_{S} \cdot M_{gas} (t)\, ,
\end{equation}
where $\nu_{S}=\alpha_{S} \sigma_g^3/(M_{lum}G)$, for the slow mode.
For the jet-induced mode we have:
\begin{equation}
\psi (t)= \nu_{J} \cdot M_{gas} (t)\, ,
\end{equation}
where $\nu_{J}=\alpha_{J} \sigma_g^3/(M_{lum}G)$.
{ Interestingly, it can be seen that we can rewrite the above equations as 
$\nu_{S,J}\simeq \alpha_{S,J}/t_{dyn}$, namely the SF time-scale
is proportional to the dynamical time of the system, and the proportionality constant
is given by either $\alpha_{S}$ or $\alpha_{J}$. In the former case the SF time-scale
will be only a (small) fraction of $t_{dyn}$, whereas in the latter the SF can proceed
on a dynamical time-scale. We can estimate the boosting factor in the SF due to the positive feedback simply as $\alpha_{J}/\alpha_{S} \sim$ 5 for the most
massive galaxies, but it can reach values as high as 50-100 for smaller objects as we will see in Sec. 3.
We also note  that in both (the slow and the jet-triggered) cases $\nu$ scales as the stellar velocity dispersion $\sigma$: therefore a downsizing
trend is built-in in either cases.}
Silk (2005) also provides a possible demarcation between these two modes for the star formation efficiency
at  around $M_\ast \sim 3\times 
10^{10}\rm M_\odot $, resembling a trend seen in the SDSS data (Kauffmann et al. 2003).
We will show that this distinction seems to be supported by our models.

\subsection{Potential energy of the gas}

{ A key quantity of the model is the potential energy of the gas,
which gives the minimum energy to be achieved in order for the 
galactic wind to develop (this also corresponds also to the time
at which we halt the SF; for details see PM04 and Pipino et al. 2002).
Following Martinelli et al. (1998), we evaluated the potential  energy of the gas as

\begin{equation}
\Omega (t) = \int_{0}^{R} d L(R)\, ,
\end{equation}
where $d L(R)$ is the work required 
to carry a quantity $dm= 4\pi R^2 \rho_{gas}(t) dR$
of mass at a radius $R$ out to infinity
(Martinelli et al., 1998). The baryonic matter (i.e. star plus gas)
is assumed to follow the distribution (Jaffe, 1983):

\begin{equation}
F_l (r)\propto {r/r_o \over 1+r/r_o }\, , 
\end{equation}
where $r_o = R_{eff}/0.763$.

We assume that the dark matter is distributed in a diffuse halo
ten times more massive than the baryonic component of the galaxy
with a scale radius $R_{dark}= 10 R_{eff}$ (Matteucci 1992), where $R_{eff}$
is the effective radius. In particular, we make use of Eq. 1 and 16 by Matteucci \& Tornambe'
(1987) in order to empirically estimate $R_{eff}$ in a given mass bin.} The effective radius $R_{eff}$ that we
refer to is the final value, however it undergoes negligible evolution 
after the collapse is almost over (this occurs roughly at time $\tau$).
{ The DM profile is taken from Bertin et al. (1992).}
We estimate the gas velocity dispersion as $\sigma_g^2= \Omega (t)/3 M_{gas}(t)$.
It is a generic function of time because the gas is subject to both the dark matter potential (assumed
to be fixed, following PM04) and the gravity due to the baryonic matter (which
changes in time)

\subsection{Supernova feedback}

One of the fundamental points upon which our model is based,
is the detailed calculation of the SN explosion rates.
For type Ia SNe, we assume a progenitor model 
made of a C-O white dwarf plus a red giant 
(Greggio $\&$ Renzini, 1983; Matteucci $\&$ Greggio, 1986). 
The predicted type Ia SN explosion rate is constrained
to reproduce the present day observed value (Cappellaro et al., 1999)
We adopt two different recipes for SNIa and SNII, respectively.

{ SNe Ia are allowed to transfer all of their initial blast wave energy,
$10^{51}\rm erg$.
The reason for this extremely efficient energy transfer 
resides in the fact that radiative 
losses from SNIa are likely to be negligible, 
since
their explosions occur in a medium already heated by SNII (Recchi et al. 2001).
On the other hand, for SNe II which explode first in a cold and dense 
medium we allow for the cooling to be 
quite efficient. In particular, } for SNII we assume that the evolution of the energy in the 'snow-plow'
phase is regulated by the Cioffi et al. (1988) cooling time.

Since we use a chemical evolution code which 
adopts the same formulation for the feedback as in Pipino et al. (2002, to which we refer the reader),
we consider a $\sim$ 20\% mean efficiency in energy transfer as a  
representative value also for the model galaxies presented in this paper.
We still define the time when the galactic wind occurs ($t_{gw}$) as the 
time at which the energy input by supernovae exceeds the gas binding
energy (see section above and Pipino et al. 2002).
The wind carries out the residual gas from the galaxies, thus inhibiting further star formation.

\subsection{Infall}

We recall that the  main novelty of the PM04 paper relative to our previous discussions (Matteucci et al., 1998; Martinelli et al., 1998;
Pipino et al., 2002) is that we simulate the creation of the spheroid as
due to the collapse of either a big gas cloud or several smaller gas lumps. 
The infall makes the star formation rate
start from a smaller value than in the closed box case,
reach a maximum and, then,  decrease when the star formation process becomes dominant.
This treatment is certainly more realistic since it includes the simulation of a gas collapse. 

Furthermore, in a galactic formation scenario in which
the galaxies are assembled starting from gas clouds falling into the galactic potential well,
the more massive objects have a higher probability (i.e. a higher cross-section)
of capturing gas clouds than the less massive systems, thus completing their assembly on a faster timescale
(Ferreras $\&$ Silk, 2003) . 
This mechanism accounts for the anti-hierarchical behaviour of SMBH,
because smaller galaxies accrete gas at a lower rate with respect to more massive
objects, thus having a smaller reservoir with which to feed the SMBH.

The infall term is present on the right-hand side of the equation of  chemical 
evolution, and the adopted expression is:
\begin{equation}
({d G_i (t) \over d t})_{infall}= X_{i,infall} C e^{-{t \over \tau}}\, ,
\end{equation}
where $X_{i,infall}$ describes the chemical composition of the accreted gas, assumed to be primordial.
$C$ is a constant obtained by integrating the infall law over time and
requiring that $\sim 90\%$ of the initial gas has been accreted at $t_{gw}$ (in fact,
we halt the infall of the gas at the occurrence of the galactic wind).
Finally, $\tau$ is the infall time-scale.

\subsection{Stellar Yields}

We follow in detail the evolution of 21 chemical elements, for which 
we need to adopt specific prescriptions for stellar nucleosynthesis.
In particular, our nucleosynthesis prescriptions are:
\begin{enumerate}
\item
For single low and intermediate mass stars ($0.8 \le M/M_{\odot} \le 8$) we make use of the yields 
of  van den Hoek $\&$ Groenewegen (1997) as a function of metallicity. 
\item
We use the yields by Nomoto et al. (1997) for SNIa which are assumed to originate from C-O white dwarfs
in binary systems which accrete  material from a companion (the secondary) and reach the Chandrasekar 
mass and explode via C-deflagration . 
\item
Finally, for massive stars ($M >8 M_{\odot}$) we adopt the yields of  Thielemann et al. (1996, TNH96)
which refer to the solar chemical composition.
\end{enumerate}

\section{Results and discussion}

We run models for elliptical galaxies in the baryonic mass range 
$10^{10}-10^{12}M_{\odot}$.
$M_{lum}$ is the 'nominal' mass of the object, i.e. the mass of the initial gas cloud (we recall that we normalize
the infall law between $t=0$ and $t\sim t_{gw}$). The mass in stars at the present time
is $\sim 0.2-0.4$ $M_{lum}$ for all the models and the velocity dispersion $\sigma$ is evaluated
from the relation $M=4.65\cdot 10^5\, \sigma^2\, R_{eff}\, M_{\odot}$ (Burstein et al., 1997). 

The models are:
\begin{enumerate}
\item Model I:  PM04's best model: Salpeter IMF, $\tau$ is decreasing with galactic mass, whereas $\nu$ increases (see Table 1).
\item Model II: Salpeter IMF, $\tau$ decreasing with galactic mass; $\nu$ self-consistently calculated according to Silk (2005) slow SF mode (see Table 1).
\item Model III: Salpeter IMF, $\tau$ decreasing with galactic mass; $\nu$ self-consistently calculated according to Silk (2005) SMBH-triggered SF mode (see Table 1).
\item Model IV: a \emph{hybrid} case between Model II - for low galactic masses - and Model III - for the high
mass end (see Table 1).
\end{enumerate}

The basic features of the above mentioned models are highlighted in Table 1,
where the input luminous mass, effective radius, SF efficiency and infall timescale
are listed in columns 1-4, respectively, whereas the output galactic wind
time-scale and mass-weighted abundance ratios in stars are presented in columns 5-7.

\begin{table*}
\centering
\begin{minipage}{120mm}
\scriptsize
\begin{flushleft}
\caption[]{Summary of model properties}
\begin{tabular}{l|llllllll}
\hline
\hline
$M_{lum}$ 	&$R_{eff}$ &  $\nu$ 	    & $\tau$& $t_{gw}$ & $[<Fe/H>]$ &$[<Mg/Fe>]$\\
({$M_{\odot}$}) & ({kpc})  &  ({$Gyr^{-1}$})& (Gyr)& (Gyr)   &        &              \\
\hline
Model I (PM04)\\
\hline
$10^{10}$       & 1        & 3  ($\nu_{PM04}$)             & 0.5  & 1.30   &  -0.10  & 0.15	     \\ 
$10^{11}$       & 3        & 10  ($\nu_{PM04}$)             & 0.4   &  0.55  & -0.06  & 0.21      \\ 
$10^{12}$       & 10       & 22  ($\nu_{PM04}$)            &0.2    &   0.44  & 0.06  & 0.33      \\
$10^{12}$b       & 10       & 44 (2x$\nu_{PM04}$) & 0.2   &  0.41  & 0.10  & 0.34     \\
\hline
Model II \\
\hline
$10^{10}$       & 1	   & $\nu_S$         & 1.5    & 0.94    & -0.18  &  0.15     \\
$10^{11}$       & 3        & $\nu_S$         & 0.4   &  0.66  & -0.04  & 0.26     \\
$10^{12}$       & 10       & $\nu_S$         & 0.2   & 0.56  & 0.08  & 0.29    \\
\hline
Model III\\
\hline
$10^{10}$       & 1	   & $\nu_J$         & 0.5    &  1.40   & -0.45  & 0.57     \\
$10^{11}$       & 3        & $\nu_J$         & 0.4   &  0.71  & -0.27  & 0.48     \\
$10^{12}$       & 10       & $\nu_J$         & 0.2   &  0.33  & -0.04  & 0.35     \\
\hline
Model IV (hybrid) plus 10\% SN eff. \\
\hline
$10^{10}$       & 1	   & $\nu_S$         & 1.5    & 1.00    & -0.05  &  0.13     \\
$10^{11}$       & 3        & $\nu_S$         & 0.4   &  0.64  & 0.00  & 0.20     \\
$10^{12}$       & 10       & $\nu_J$         & 0.2   &  0.50  & 0.21  & 0.32     \\
\hline
\end{tabular}
\end{flushleft}
\end{minipage}
\end{table*}

We start the analysis by comparing the massive objects predicted by model III with the PM04 best model (model I).
The similarity is striking, even if $\nu_J > \nu_{PM04}$, more than in the case in which
we simply double the SF efficiency (Model I, $10^{12}$b). This happens because of the
linear relation between $\nu_J$ and time. In fact, as  time proceeds, the gas
mass decreases mainly due to SF, whereas $\Omega$ increases because
of the low- and intermediate-mass stars formed.

\begin{figure}
%\epsscale{1.5}
%\figurenum{4}
\epsfig{file=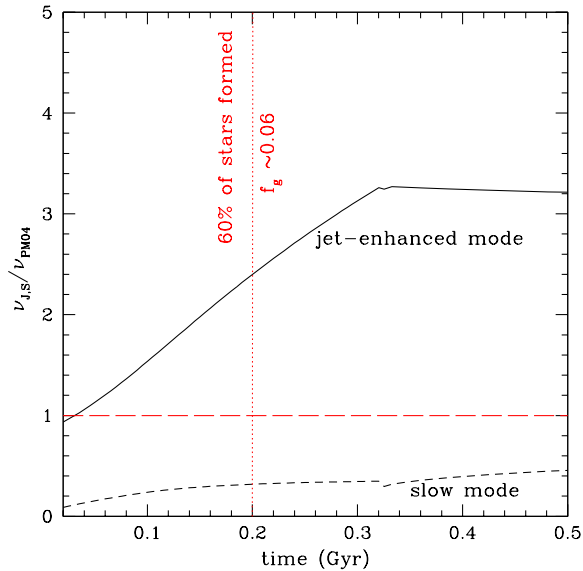,height=8cm,width=8cm}
\caption{ The time evolution of the SF efficiency ($\nu$) normalized to the PM04 value (see table 1) in the case of 
a $10^{12}M_{\odot}$ object. Solid line: jet-enhanced formation mode ($\nu_J$, Model III);
dashed line: slow mode ($\nu_S$, Model II).
Dashed horizontal line: $\nu_{J,S} = \nu_{PM04}$. 
\label{fig1}}
\end{figure}

From fig.~\ref{fig1} we find that
$\nu_J$ is only higher than $2\times \nu_{PM04}$ at a late stage.
For most of the galactic active evolution, however,  $\nu_J \sim \nu_{PM04}$, therefore
the global effect on the SN explosion rate and the resulting galactic wind are similar.
The slow mode (Model II), instead, has an efficiency $\nu_S < \nu_{PM04}$ ({ but still within a factor
of 2-3}) for most of the galactic
evolution. This produces a lower $\alpha$-enhancement with respect to the average
value inferred from observations for the given galactic mass. 

\begin{figure}
%\epsscale{1.5}
%\figurenum{4}
\epsfig{file=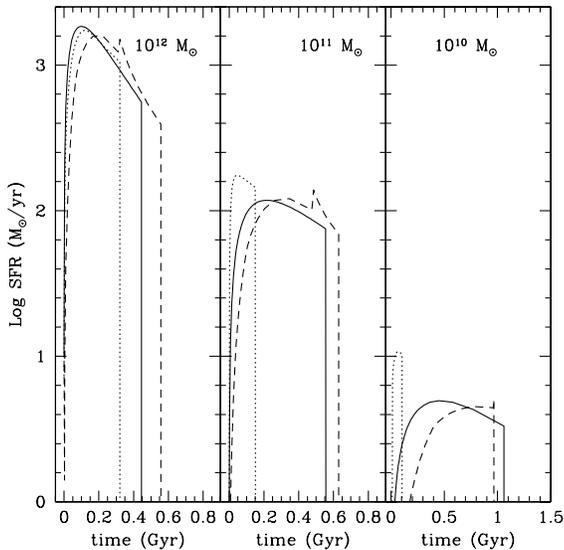,height=8cm,width=8cm}
\caption{ The time evolution of the SF rate for different mass models as
reported in each panel.
Solid line: PM04 model ($\nu_{PM04}$, Model I);
dashed line: slow mode ($\nu_S$, Model II);
dotted line: jet-enhanced formation mode ($\nu_J$, Model III).
\label{fig2new}}
\end{figure}

{ With the help of Fig.~\ref{fig2new} we can see the effect of the two modes
on the SF histories for three different masses. The solid line refers to the standard SF history of the PM04 model,
which, in a sense, is a template that we want to reproduce with the more physically grounded
recipe for evaluating $\nu$. The new models are presented by dashed (Model II) and dotted (Model III) lines, respectively.
Again, we stress that the fact that $\nu_J \sim \nu_{PM04}$ for the most massive galaxy
leads to a SF history which is very similar to the one predicted by PM04.}

{ At the low mass end, such a model implies $\nu_J >> \nu_{PM04}$, hence
too large an amount of SNII which trigger too early a wind with respect to the fiducial PM04 case. The
result is an average [Mg/Fe] higher than expected from observations at that given mass.}
The  slow SF mode (Model II) seems to be more appropriate in this mass range.

The analysis of an intermediate-size galaxy ($M_{lum}= 10^{11}M_{\odot}$)
helps in understanding at which mass the SF process switches from the slow
to the SMBH-induced mode.
We find  better agreement with { the fiducial PM04 SF history} when the model with $M_{lum}= 10^{11}M_{\odot}$
undergoes a slow SF mode, { whereas the jet-triggered case leads to a SF history lasting less than 0.2 Gyr.} 
The stellar mass of this model after $\sim$ 10 Gyr
of passive evolution is $\sim 0.4 \times 10^{11}M_{\odot}$.
{ Even if at a first glance of Fig.~\ref{fig2new} it may seem that the \emph{downsizing}
trend with galactic mass, namely that the more massive galaxies have a higher peak
value for the SF rate than the lower mass ones, is fulfilled by every SF recipe, we warn the reader that this does not suffice
to reproduce the MFMR. At the same time the duration of the star formation must decrease with mass. 
Moreover, since smaller galaxies are younger than the most massive ones (e.g. Thomas
et al. 2005), we expect the peak in the SF rate, namely the epoch at which the majority
of the stars in a given galaxy form, to shift to later times.}

\begin{figure}
%\epsscale{1.5}
%\figurenum{4}
\epsfig{file=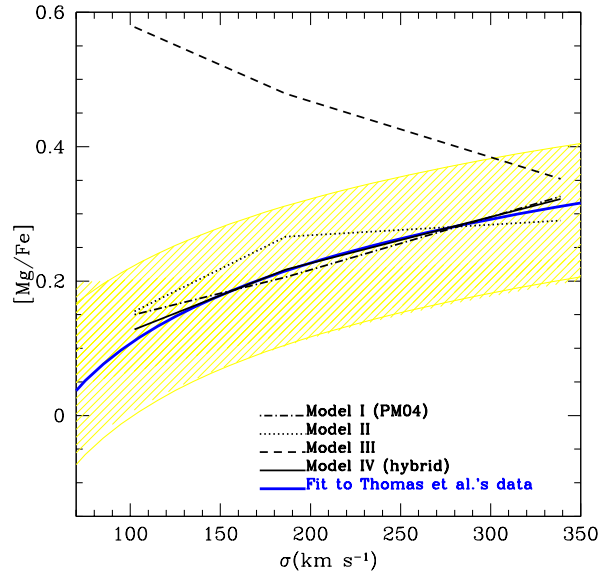,height=8cm,width=8cm}
\caption{[Mg/Fe] as a function of galactic velocity dispersion predicted by Model I (dot-dashed),
II (dashed), III (dotted) and by the best combination of these two (see text - solid) compared to the
data by Thomas et al. (2008). The formal linear regression to Thomas et al. (2008) is
given as a thick solid line that spans the entire range in $\sigma$.
\label{fig2}}
\end{figure}

{ These findings become more clear if we show}
the MFMR predicted by different models plotted against a
set of data { (Fig.~\ref{fig2}). Although several large samples
are now available (e.g. Nelan et al. 2006, Smith et al. 2006, Graves et al. 2007), almost
all agree on the typical slope and zero-point of the MFMR; therefore we make
use of only one of such datasets, namely a sample of ellipticals drawn out
of the SDSS, whose line-strength indices have been} analysed and transformed into [$\alpha$/Fe] ratios by Thomas et al. (2008).
Not surprisingly, Model I (dash-dotted line) fits the observed relation very well, being also the PM04 best model.
As expected from the above discussion, Model II (dotted line) gives a flattening at high masses,
whereas Model III (dashed line) predicts galaxies with a high overall $\alpha$ enhancement and a decreasing
trend with mass, in disagreement
with observations.

{In order to make a more quantitative comparison, we note
that a linear regression fit of Thomas et al. (2008)'s data returns $[\alpha/Fe]=-0.56+0.34\, log \sigma$ (Fig.~\ref{fig2}:thick solid
line that covers the entire range in $\sigma$), with
an \emph{intrinsic scatter} in the relation of $\sim$0.1 dex.
In the case of model I, we predict $[Mg/Fe]=-0.51+0.33\, log \sigma$, therefore the agreement is
remarkable. Model II, gives a slightly flatter trend, $[Mg/Fe]=-0.35+0.25\, log \sigma$
{ in the whole mass range, but basically a null slope at the high mass end}; nonetheless
its predictions are discrepant by only one standard deviation\footnote{We assume that
the\emph{intrinsic scatter} quoted by Thomas et al. (2008) corresponds to
one standard deviation, although they do not report any formal error on both
the intercept and the slope obtained from the fit to they data.}. Finally, Model III is
clearly ruled out, because the slope of the MFMR is negative and the zero point
is offset by more than an order of magnitude from the data; in fact the formal linear regression
returns $[Mg/Fe]=1.39 - 0.41\, log \sigma$.}

\begin{figure}
%\epsscale{1.5}
%\figurenum{4}
\epsfig{file=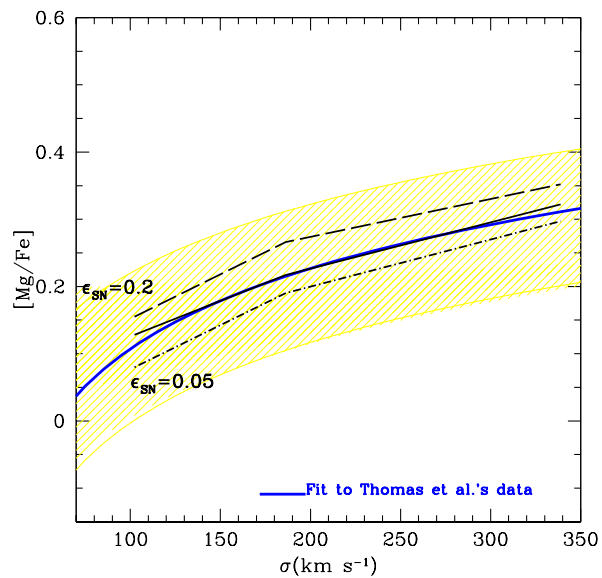,height=8cm,width=8cm}
\caption{[Mg/Fe] as a function of galactic velocity dispersion predicted by Model IV (solid)
for other three supernova feedback efficiencies, compared to
data by Thomas et al. (2008).
\label{fig3}}
\end{figure}

We now let only the most massive galaxies form through a SMBH-induced SF episode,
whereas the lower mass ones are allowed to form only via the slow mode.
From the entries of Table 1, however, 
we notice that both Model II and III galaxies populate the upper half of the MFMR
therefore we are probing the shortest possible SF time-scales,
{ as it can be seen by a close inspection of Fig.~\ref{fig2new}}.
In order to have a better fit to the mean trend, we let the SF timescale
be slightly longer by lowering the SN ejection efficiency to the 10\%.
{ Basically, since $\nu_J$ is slightly higher than $\nu_{PM04}$, the time
at which the galactic wind sets in ($t_{gw}$) occurs earlier, because
the a suitable number of SN explosions occurs on a shorter timescale. An easy
way to compensate for this effect - without changing the stellar metal production,
the SFR and, hence, the SN rate itself- is to lower the mean energy input to the interstellar
medium by SNe. In particular, this is done by halving the energy input from SNIa, since SNII
already undergo cooling and contributed only to a few percent of the total budget (see Sec. 2.4).}
In this case we get the so-called \emph{hybrid} case represented by the solid line (Model IV in Table 1, solid
line in Figs.~\ref{fig2new} and ~\ref{fig3}).
It reproduces fairly well the observed trend, having practically the same
values of the PM04 best model at a given galactic mass.{ In this case, in fact, 
we predict $[Mg/Fe]=-0.57+0.35\, log \sigma$, in excellent agreement with Thomas et al.'s data.}
This last \emph{hybrid} case works well in reproducing the MFMR,
whose scatter can be explained by variations of \emph{local} properties
from galaxy to galaxy.
In particular, we also show in fig.~\ref{fig3} the cases in which we set
the total SN efficiency to the PM04 default value (i.e. 20\%
on the average) and to 5\%, respectively.
{ Interestingly, the best match is obtained by lowering the SN efficiency at variance
with many theoretical works which, instead, require extremely high values in order to, e.g., trigger
a wind (see Pipino et al., 2002 for discussion and references).
We note that values smaller than 5\% hardly allow the galactic wind to develop, 
whereas fractions much higher that 20\% unbind the gas too early, leading to a situation
which will be more similar to a disruption of the galaxies rather than a galactic wind.
In the \emph{allowed} region, instead, the SN efficiency can be regarded as a free parameter.
In this regime, the SFR - which consumes gas and create massive stars which
will explode as SNII - set the conditions for the wind
to happen around 0.5 Gyr, whereas the contribution of SNIa is important in determining
the exact occurrence of the wind. 
An advantage of the new formulation for $\nu$ is that it helps in reducing the SNIa transfer
efficiency from the somehow unphysical value of 100\% down to a more sensible percentage.
We also deem implausible to entirely explain the MFMR by means a change in the SN efficiency - increasing
as a function of mass - at a fixed $\nu$.}
{ The effect of changes in other parameters, such as the infall timescale and the IMF have
already been addressed in Thomas et al. (1999) and PM04, whereas the degeneracy between star
formation efficiency and infall timescale was thoroughly discussed in Ferreras \& Silk (2003); therefore
we do not repeat here their analysis.
Small variations in the above quantities can easily make the model galaxy properties
span the observed range and recover the scatter. The important conclusion
that we draw from our phenomenological approach is that both the
infall and the star formation time-scales have to be much shorter in ellipticals
than in spirals.}

{ In principle, different prescriptions for the nucleosynthesis can affect the predicted quantities.
As shown by PM04 (see also Thomas et al. 1999), a change in the yields for massive stars
can produce a 0.2--0.3 dex difference in the final stellar [Mg/Fe]. However,
when one makes use of recipes for the nucleosynthesis which have been constrained
in the solar vicinity, the differences are smaller (see discussion in PM04).

Also, the type Ia progenitors and the mean delay in their occurrence since
the episode of star formation are debated in the literature. On the basis of observational arguments, 
Mannucci et al. (2005, 2006) recently suggested  
that there is a bimodal distribution of delay times for the explosion of Type Ia SNe. 
In particular, a percentage from 35 to 50\% of 
the total Type Ia SNe should be composed by systems with lifetimes as 
short as $10^{8}$ years, whereas the rest should arise from smaller mass progenitors 
with a much broader distribution of lifetimes. Matteucci et al. (2006) tested Mannucci et al.'s 
hypothesis in models of chemical evolution of 
galaxies of different morphological type: ellipticals, spirals and irregulars. 
They showed that this proposed scenario is compatible also 
with the main chemical properties of galaxies.
For ellipticals the differences are less noticeable than in other morphological types, since 
they must have evolved very fast and at a very high redshift. 
The differences produced in the [Mg/Fe] ratio in the stars with respect to the 
results of PM04 are negligible. }

{ In summary, the slow mode works well at low and intermediate galactic masses where we want to have [Mg/Fe] $<$ 0.2 dex on average, 
namely in a situation resembling more a low-rate star forming disc  ($\nu \sim 1$) rather than a massive elliptical.
On the other hand, the extremely short SF rates in Fig.~\ref{fig2new} for the lower mass spheroids, with
the consequent [Mg/Fe] ratios 0.4 dex off the observed MFMR, tell us that these galaxies
have to form stars on a timescale much longer than the dynamical time.
Therefore, the jet-triggered mode in not suitable in the low mass regime.
Indeed Silk (2005) already finds that the transition to jet-triggered mode occurs only at stellar masses larger than $3\times 10^{10}M_{\odot}$, value
which is close to the final stellar mass of our intermediate-mass case.
As a matter of fact, the jet-triggered mode proves to be reasonable only for the largest masses. This, in a sense, is reassuring, because
the most massive objects are the most difficult to explain in any scenario of galaxy formation.
Therefore they need some extreme recipe. 

We also find remarkable the fact that, if we estimate the boosting factor in the SF due to the positive 
feedback simply as $\alpha_{J}/\alpha_{S}$, we have only a factor of 5 (as shown
in Fig.~\ref{fig1}) for the most
massive galaxies. In other words, as shown in Fig.~\ref{fig2new}, the SF rate predicted
by the slow-mode for the most massive galaxy does not differ much from the fiducial
behaviour of PM04. The reason for the similarity between the two modes
at high masses is linked to the fact that $\nu_s \sim \sigma_g^4$, whereas $\nu_s \sim \sigma_g^3$.
On the other hand, a factor of 5 increase in the star formation efficiency suffices to create differences in the
final [Mg/Fe]; for instance the difference between $\nu_{PM04}$ for the lowest and the intermediate mass case is $\sim$3,
whereas the difference with respect to the moss massive case amount to a factor of $\sim$7. 
We prefer the "hybrid" case because it gives the best match with the observed slope
in the MFMR, especially in the high galactic mass regime.
In this sense the SMBH positive feedback shapes the MFMR, but only at the high
mass end. 
The important novelty of our approach to the modelisation
of the SF efficiency is that in both (the slow and the jet-triggered) cases $\nu$ scales as $\sigma$. This is
what creates the \emph{downsizing} behaviour.
Changes in other model parameters - we explored the SN feedback efficiency, but stellar yields
and IMF can have a role - affect mainly the zero-point of the MFMR and help us to achieve a \emph{best match} of the mean
observational trend, but cannot create the slope in the MFMR\footnote{Unless an unphysical variation of the stellar yields
or a flattening of the IMF with galactic mass are invoked.}. 
Unfortunately, the degree of degeneracy between model parameters is still quite high, and this hamper
us from deriving more quantitative conclusions.}

\section{Conclusions}

In this paper we have  analysed the role of star formation  efficiency
in the creation of the mass-[Mg/Fe] relation (MFMR). 
We started from the heuristic approach of PM04,
who required the SF efficiency to increase as a function
of galactic mass, and  we showed that the data can be explained 
 by implementing
a physically motivated value for the SF efficiency parameter $\nu$.

To explain the higher star formation efficiency in the most massive galaxies,
we appeal to SMBH-triggered SF. Following the arguments in Silk (2005), we argue that
a short ($10^6-10^7$ yr) super-Eddington phase can 
provide the accelerated triggering of associated star formation.
The SMBH grows mostly in the initial super-Eddington phase while 
most of the spheroid stars grow during the succeeding Eddington phase, until
the SN-driven wind quenches  SF. The quenching is accomplished by the multiplicative factor of SN energy input
 induced by AGN outflows,
and results in the usual SMBH mass-spheroid velocity dispersion (Magorrian) relation. 
According to our models, the galaxy is fully assembled on a timescale of 0.3-0.5 Gyr.
This timescale is long enough, however, to allow the SMBH to complete its growth
in order to reproduce the Magorrian  relation.

In low- and intermediate-mass ETGs, instead, the SF efficiency
occurs via a \emph{slow} mode which is related to the Schmidt-Kennicutt law.
In fact,  supernova-induced feedback controls galaxy formation 
by rendering star formation slow and inefficient (relative to star formation in the
most massive galaxies). By comparing the trend in the MFMR predicted by means of these two recipes for SF, 
we find that there must be a demarcation between these two modes at
$10^{10}\rm M_\odot $, resembling a trend seen in the SDSS data 
(Kauffmann et al. 2003).
The observational scatter in the MFMR can be entirely explained as being
intrinsic. Local effects, such as variations in the SNe feedback efficiency of order a factor
of 2 with respect to the best model case, can  induce or delay 
the occurrence of the galactic wind and thus  contribute to setting the final value
for  [Mg/Fe].

\section*{Acknowledgments} 
A.P. thanks N.Nesvadba and M.Lattanzi for enlightening discussions.
The anonymous referee is acknowledged for comments that improved the
quality of the paper.

\label{lastpage}

\end{document}